\documentclass[draft,grl]{agutex}

 \usepackage{graphicx}
  \setkeys{Gin}{draft=false}
\authorrunninghead{LOI ET AL.}

\titlerunninghead{Real-time imaging of density ducts}

\authoraddr{Corresponding author: S.~T.~Loi,
Sydney Institute for Astronomy, School of Physics, The University of Sydney, NSW 2006, Australia. (sloi5113@uni.sydney.edu.au)}

\begin{document}

%
%

\title{Real-time imaging of density ducts between the plasmasphere and ionosphere}
%
%

%
%









\def\USydney{$^{1}$}
\def\CAASTRO{$^{2}$}
\def\Newcastle{$^{3}$}
\def\Haystack{$^{4}$}
\def\Curtin{$^{5}$}
\def\ANU{$^{6}$}
\def\ASTRON{$^{7}$}
\def\CASS{$^{8}$}
\def\UToronto{$^{9}$}
\def\MIT{$^{10}$}
\def\UWisc{$^{11}$}
\def\SKASA{$^{12}$}
\def\CfA{$^{13}$}
\def\Rhodes{$^{14}$}
\def\ASU{$^{15}$}
\def\RRI{$^{16}$}
\def\UW{$^{17}$}
\def\Victoria{$^{18}$}
\def\Tata{$^{19}$}
\def\UMelbourne{$^{20}$}

\authors{
Shyeh Tjing Loi\USydney$^,$\CAASTRO, 
Tara Murphy\USydney$^,$\CAASTRO, 
Iver H.~Cairns\USydney,
Frederick W.~Menk\Newcastle,
Colin L.~Waters\Newcastle,
Philip J.~Erickson\Haystack,
Cathryn M.~Trott\CAASTRO$^,$\Curtin,
Natasha Hurley-Walker\Curtin,
John Morgan\Curtin,
Emil Lenc\USydney$^,$\CAASTRO,
Andr\'{e} R.~Offringa\CAASTRO$^,$\ANU$^,$\ASTRON,
Martin E.~Bell\CAASTRO$^,$\CASS,
Ronald D.~Ekers\CASS,
B.~M.~Gaensler\USydney$^,$\CAASTRO$^,$\UToronto,
Colin J.~Lonsdale\Haystack,
Lu Feng\MIT,
Paul J.~Hancock\CAASTRO$^,$\Curtin,
David L.~Kaplan\UWisc,
G.~Bernardi\SKASA$^,$\CfA$^,$\Rhodes,
J.~D.~Bowman\ASU,
F.~Briggs\CAASTRO$^,$\ANU,
R.~J.~Cappallo\Haystack,
A.~A.~Deshpande\RRI,
L.~J.~Greenhill\CfA,
B.~J.~Hazelton\UW,
M.~Johnston-Hollitt\Victoria,
S.~R.~McWhirter\Haystack,
D.~A.~Mitchell\CAASTRO$^,$\CASS,
M.~F.~Morales\UW,
E.~Morgan\MIT,
D.~Oberoi\Tata,
S.~M.~Ord\CAASTRO$^,$\Curtin,
T.~Prabu\RRI,
N.~Udaya~Shankar\RRI,
K.~S.~Srivani\RRI,
R.~Subrahmanyan\CAASTRO$^,$\RRI,
S.~J.~Tingay\CAASTRO$^,$\Curtin,
R.~B.~Wayth\CAASTRO$^,$\Curtin, 
R.~L.~Webster\CAASTRO$^,$\UMelbourne, 
A.~Williams\Curtin,
C.~L.~Williams\MIT
}

$^{1}$Sydney Institute for Astronomy, School of Physics, The University of Sydney, NSW 2006, Australia\\
$^{2}$ARC Centre of Excellence for All-sky Astrophysics (CAASTRO)\\
$^{3}$School of Mathematical and Physical Sciences, The University of Newcastle, Callaghan, NSW 2308, Australia\\
$^{4}$MIT Haystack Observatory, Westford, MA 01886, USA\\
$^{5}$International Centre for Radio Astronomy Research, Curtin University, Bentley, WA 6102, Australia\\
$^{6}$Research School of Astronomy and Astrophysics, Australian National University, Canberra, ACT 2611, Australia\\
$^{7}$Netherlands Institute for Radio Astronomy (ASTRON), Postbus 2, 7990 AA Dwingeloo, The Netherlands\\
$^{8}$CSIRO Astronomy and Space Science (CASS), PO Box 76, Epping, NSW 1710, Australia\\
$^{9}$Dunlap Institute for Astronomy and Astrophysics, The University of Toronto, ON M5S 3H4, Canada\\
$^{10}$Kavli Institute for Astrophysics and Space Research, Massachusetts Institute of Technology, Cambridge, MA 02139, USA\\
$^{11}$Department of Physics, University of Wisconsin--Milwaukee, Milwaukee, WI 53201, USA\\
$^{12}$Square Kilometre Array South Africa (SKA SA), Cape Town 7405, South Africa\\
$^{13}$Harvard-Smithsonian Center for Astrophysics, Cambridge, MA 02138, USA\\
$^{14}$Department of Physics and Electronics, Rhodes University, PO Box 94, Grahamstown, 6140, South Africa\\
$^{15}$School of Earth and Space Exploration, Arizona State University, Tempe, AZ 85287, USA\\
$^{16}$Raman Research Institute, Bangalore 560080, India\\
$^{17}$Department of Physics, University of Washington, Seattle, WA 98195, USA\\
$^{18}$School of Chemical \& Physical Sciences, Victoria University of Wellington, Wellington 6140, New Zealand\\
$^{19}$National Centre for Radio Astrophysics, Tata Institute for Fundamental Research, Pune 411007, India\\
$^{20}$School of Physics, The University of Melbourne, Parkville, VIC 3010, Australia

%
%


\begin{abstract}
Ionization of the Earth's atmosphere by sunlight forms a complex, multi-layered plasma environment within the Earth's magnetosphere, the innermost layers being the ionosphere and plasmasphere. The plasmasphere is believed to be embedded with cylindrical density structures (ducts) aligned along the Earth's magnetic field, but direct evidence for these remains scarce. Here we report the first direct wide-angle observation of an extensive array of field-aligned ducts bridging the upper ionosphere and inner plasmasphere, using a novel ground-based imaging technique. We establish their heights and motions by feature-tracking and parallax analysis. The structures are strikingly organized, appearing as regularly-spaced, alternating tubes of overdensities and underdensities strongly aligned with the Earth's magnetic field. These findings represent the first direct visual evidence for the existence of such structures.
\end{abstract}

%
%

%

\begin{article}

%
%

\section{Introduction}
The plasmasphere is a toroidal region within the Earth's magnetosphere that is filled with plasma from the ionosphere, which is a photoionized layer near the surface of the Earth \citep{Goldstein2006}. The existence of field-aligned ducts in the plasmasphere is widely accepted as the explanation for ground-based detections of whistler-mode waves, which may be guided along such ducts \citep{Ohta1996,Bakharev2010}. Wave-particle interactions that accelerate and precipitate particles into the atmosphere occur efficiently within these ducts, so their existence aids the removal of energetic particles from the magnetosphere and thereby influences global magnetospheric-ionospheric coupling and energetics \citep{Sonwalkar2006}. Duct properties have been studied mainly through indirect, ground-based observations of whistlers from both natural sources (lightning) and man-made VLF transmitters \citep{Singh1998,Ganguli2000,Clilverd2003}. Since the guiding structures cannot be detected independently of the whistlers, observations that use lightning-generated whistlers are necessarily intermittent, and those that use man-made whistlers are spatially restricted. Direct detections of plasmaspheric ducts have been achieved by satellite in-situ measurements \citep{Sonwalkar1994}, but these and also indirect satellite-based measurements such as sounding  \citep{Darrouzet2009,Yearby2011,Karpachev2014} are heavily restricted by spacecraft orbital patterns. Duct morphologies have thus been difficult to establish.

The Murchison Widefield Array \citep[MWA;][]{Lonsdale2009,Tingay2013} is a low-frequency (80--300\,MHz) radio telescope located in Western Australia at $26^\circ 42' 12''$S, $116^\circ 40' 15''$E (geographic), and a geomagnetic latitude of 38.6$^\circ$S (McIlwain $L$ parameter \citep{McIlwain1961} of 1.6). Although chiefly intended for studying the cosmos, it is also an exquisitely sensitive probe of the near-Earth plasma. Novel design concepts include its wide field-of-view (FoV) and high-cadence imaging capability. A multitude of unresolved celestial radio sources (mostly radio galaxies and quasars) back-illuminate the plasma, allowing one to probe fluctuations at high spatial completeness by measuring the angular distribution of source refractive shifts. Since interferometers measure angular positions using baseline phase differences, they are insensitive to the constant offset component of the electron column density (``total electron content'', or TEC), which adds an equal phase to all antennas and cancels out. Rather, the refraction-induced angular shift $\Delta \theta$ of a radio source depends on the transverse gradient $\nabla_\perp \mathrm{TEC}$ in the TEC towards the source, and is given by \citep{Thompson2001}
\begin{equation}
  \Delta \theta = -\frac{1}{8\pi^2} \frac{e^2}{\epsilon_0 m_e} \frac{1}{\nu^2} \nabla_\perp \mathrm{TEC} \:. \label{eq:posshift}
\end{equation}
Here $e$ and $m_e$ are the electron charge and mass, $\epsilon_0$ the vacuum permittivity, and $\nu$ the radio observing frequency. The negative sign indicates that the direction of refraction is towards decreasing TEC.

\section{Observations and Results}
We analyzed MWA data comprising 46 snapshots, each integrated for 2\,min at a frequency of 183\,MHz over a 30.72\,MHz bandwidth, of a $30^\circ \times 50^\circ$ (N-S $\times$ E-W) patch of sky near zenith. These were recorded over a 1.5-hr interval on 2013-10-15 between 1346--1517 UTC (pre-midnight local time), in which the telescope was tracking celestial coordinates RA = 0$^\circ$, Dec = $-27^\circ$ (J2000). We measured the refractive shifts of the $\sim$1000 celestial sources visible in each $\sim$30$^\circ$-wide snapshot by fitting to the positions of the same sources seen in each snapshot and computing offsets from the time-averaged positions. In more detail, the source-finding software \textsc{Aegean} \citep{Hancock2012} was first used to locate intensity peaks (candidate radio sources) in the images. We subjected the candidate sources to a number of quality restrictions and then cross-matched the remainder with the NRAO VLA Sky Survey (NVSS) catalog \citep{Condon1998}, a published database of celestial radio sources, retaining only those candidates with counterparts in NVSS. Cross-matching was performed by identifying the nearest NVSS source to the candidate source within the search radius (3\,arcmin) and associating it to the latter.

We took the reference position of a source, uniquely identified by its NVSS catalog name, to be the average of its celestial coordinates (measured by \textsc{Aegean}) over all snapshots. Then for each snapshot, we computed the angular offset vector of each source from its reference position. The distributions of these offsets (equivalently the TEC gradient field) are shown in Fig.~\ref{fig:arrowplots} for several snapshots (see Movie S1 for an animation of the full dataset). Strikingly organized bands of arrows are visible, with a characteristic band separation of $\sim$2$^\circ$. These reflect a spatially oscillating gradient field, implying alternating bands of underdensities and overdensities. The bands are tightly aligned with the Earth's magnetic field. Taking the divergence of the vector field reveals the duct structure more prominently. This is shown in Fig.~\ref{fig:divplot} for a representative snapshot (see Movie S2 for an animation of the full dataset).

The measured offsets display a $\lambda^2$ proportionality (Fig.~\ref{fig:trends}a) consistent with Equation (\ref{eq:posshift}), where $\lambda$ is the observing wavelength. Source motions exceed measurement errors and are often larger than the width of the telescope point-spread function (119\,arcsec at 183\,MHz). Quasi-sinusoidal oscillations of radio sources are apparent; the time series for one source, with errors, is shown in Fig.~\ref{fig:trends}b. Note that celestial sources drift across the sky much faster than do the structures, which are almost stationary above the MWA. Their characteristic TEC gradient (a measure of the amplitude), obtained from the mean angular offset using Equation (\ref{eq:posshift}), increases over time and roughly doubles in 1.5\,hr (Fig.~\ref{fig:trends}c). This is comparable to the growth timescales predicted and reported for whistler ducts in the literature \citep{Singh1998, McCormick2002}.

We estimated the altitude, motion and orientation of the structures by using the MWA as a stereo camera. The MWA consists of 128 receiving elements (``tiles'') spread over an area about 3\,km across (Fig.~\ref{fig:stereo}a). By dividing the array east-west and imaging using the two groups of tiles separately, we measured a parallax shift (Fig.~\ref{fig:stereo}b) that implies a characteristic altitude of $570 \pm 40$\,km ($L = 1.8$), averaged over the interval and the FoV. This is within the $L$-range previously inferred for similar structures \citep{Jacobson1993,Hoogeveen1997}. 

A parallax analysis is only meaningful for structures that are localized in altitude. However, the MWA measures the spatial derivative of a column-integrated quantity, and so is sensitive to irregularities over a wide range of altitudes. An argument against the structures being extended in altitude, and thus for the applicability of the parallax analysis, comes from the observation that the bands are still prominent up to zenith angles of $\zeta \sim 20^\circ$. If they were extended in altitude (like vertical sheets) then oblique lines of sight would pierce multiple sheets, causing the features to blend into one another and wash out. Clear bands in the $\nabla_\perp$TEC vector field exist out to $\zeta \sim 20^\circ$, implying that the irregularities cannot be extended in altitude by more than $\cot \zeta \approx 3$ times the inter-band separation, i.e.~that their horizontal and vertical cross-sectional widths are comparable. This further supports the cylindrical duct interpretation, as opposed to vertical sheet-like structures. The contribution to the TEC from the smooth plasma above and below the irregularity layer cancels away when the horizontal spatial derivative is taken, and so the situation is observationally equivalent to a thin layer of plasma embedded with irregularities that is bounded above and below by vacuum. This supports the validity of the parallax technique, and explains why we obtained a well-defined altitude upon measurement.

Computing the parallax separately over the first and second halves of the interval establishes a downwards drift from $720 \pm 90$\,km to $470 \pm 40$\,km with time. Splitting the data spatially into northern and southern halves of the FoV reveals larger altitudes to the north. This is consistent with the steep magnetic inclination at the MWA site ($-60^\circ$), and implies that the structures stretch between $\sim$400--1000\,km within the FoV, likely extending above this. Thus they bridge the topside ionosphere and inner plasmasphere, and so the transition from a neutral- to plasma-dominated medium. These properties are consistent with the altitudes that whistler ducts are expected to occupy at night \citep{Sonwalkar2006}. The physical spacings of the tubes are then 10--50\,km, and associated TEC fluctuations between about 0.1 and 0.7\,TECU (1\,TECU = $10^{16}$\,el\,m$^{-2}$), again in agreement with inferred properties of whistler ducts \citep{Sonwalkar2006}.

\section{Discussion}
\subsection{Previous Capabilities}
While the hypothesized existence of field-aligned density ducts accounts for a large body of experimental data \citep{Ohta1996,Singh1998,Carpenter2001}, it has not been previously possible to directly verify that large, cylindrical density structures greatly extended along the Earth's magnetic field lines exist, or to track their motions precisely and continuously over regional scales. Satellite in-situ observations \citep{Sonwalkar1994} can only measure densities at single points, and cannot instantaneously probe the regional-scale density distribution. Topside sounding data and whistler spectrograms have been interpreted in terms of signal guidance by field-aligned ducts \citep{Darrouzet2009,Yearby2011}, but such methods rely on assumptions about wave propagation \citep{Singh1998}, and detections are only possible under suitable conditions. They also suffer from spatial restrictions imposed by satellite orbital paths and the placements of ground-based receivers. With a model for the Earth's magnetic field, radial motions of ducts and their $L$-shell values can be deduced from whistler spectrograms, but zonal drifts cannot be similarly established \citep{Saxton1989,Carpenter2001}. 

Claims of regularly-spaced, regional-scale, field-aligned density structures with properties similar to those seen here have been made based on observations by the Very Large Array (VLA) radio interferometer \citep{Jacobson1993,Helmboldt2012,Helmboldt2012c} and the Los Alamos plasmaspheric drift radio interferometer \citep{Jacobson1996,Hoogeveen1997}. Although these studies could measure zonal drifts, identification of the structures relied on fitting plane waves to sparsely-sampled measurements, a model whose suitability could not be independently verified. The assumption that the irregularities were field-aligned, together with a model for the Earth's magnetic field, were required to infer their altitudes. Many of these assumptions are unnecessary here. The sampling completeness of the MWA surpasses that of the VLA and Los Alamos instruments by a factor of $\sim$100, making it the first radio telescope capable of imaging the plasma. This has allowed us to visualize for the first time the regional-scale, field-aligned nature of these density structures, and to demonstrate a clear and remarkable spatial periodicity in their layout, without any prior assumptions about their morphology. Unlike these and other ground-based studies, our parallax technique to establish the altitudes requires no knowledge of the Earth's magnetic field. By removing the reliance on wave transmission for detection of the ducts, they may be monitored passively and continuously for long durations, including periods when structural distortions or imperfections may prevent them from sustaining wave propagation.

\subsection{Formation Mechanism}
Plasmaspheric density irregularities appear more often during geomagnetically disturbed periods, but the underlying formation mechanisms are not well established \citep{Hoogeveen1997,Darrouzet2009}. Geomagnetic conditions during the interval were mildly unsettled with a $K_p$ index of 2, the observations taking place in the recovery phase of a moderate storm (maximum ring current of $D_{st} = -45$\,nT near 04 UTC, 24-hr maximum $K_p$ of 4). Theoretical studies indicate that ripples and undulations in the plasmaspheric electron density may be produced by an interchange instability that preferentially occurs just after periods of strong magnetospheric convection \citep{Sazykin2004, Buzulukova2008}. Such ripples have been observed on a global scale in satellite observations \citep{Goldstein2005}. Although below the spatial scale of the grids used in those simulation studies, smaller duct-like structures such as those seen in our data may form if this mechanism continues to operate on smaller spatial scales. We note that field-aligned structures of lesser prominence and multiplicity often appear in MWA data under quieter geomagnetic conditions. They are visible on around half of all nights inspected so far, most of which (with the exception of the current dataset) were not obtained during periods of storm-time activity.

Organized azimuthal structures in the outer plasmasphere have been attributed to $\mathbf{E} \times \mathbf{B}$ convection under the influence of ultra-low frequency (ULF) standing waves \citep{Adrian2004}. Although the spatial scales of ULF waves greatly exceed those observed here, small-scale density structures can grow by flux-tube interchange under the imposition of a much larger-scale electric field \citep{Rodger1998}. The steady downward transport established above is consistent with $\mathbf{E} \times \mathbf{B}$ convection for a westward electric field of $\sim$1\,mV\,m$^{-1}$, a reasonable value for periods of substorm activity \citep{Carpenter2001}. Quasi-periodic, field-aligned structures called plasma bubbles form through a related process, where travelling ionospheric disturbances (TIDs) trigger the Rayleigh-Taylor instability \citep{Buhari2014}. However, plasma bubble formation is only viable near the equator where near-horizontal geomagnetic field lines can support the plasma against gravity, and not at mid-latitudes. The fluctuations are unlikely to be TIDs themselves, which have a dispersion relation \citep{Shiokawa2013} grossly inconsistent with the spatial and temporal properties measured here. Furthermore, TID-like fluctuations routinely appear in MWA data, and these exhibit markedly different properties from those seen here.

Spatially-varying electric fields in the plasmasphere may cause localized compressions, squeezing plasma down selected flux tubes \citep{Park1974}. Secondary ionization from particle precipitation may also produce flux tube-selective densifications. Although natural precipitation events are largely confined to the auroral zones, the powerful VLF navigation beacon at North West Cape ($\sim$400\,km north of the MWA) is known to cause electron precipitation concentrated towards the south \citep{Parrot2007}. A Fourier analysis of our data reveals the existence of structures drifting in opposite directions. Components with larger spatial frequencies are observed to drift eastward, while those with smaller spatial frequencies drift westward. We measure drift speeds within 2\% of corotation, consistent with these being in the inner plasmasphere where corotation is relatively strict \citep{Galvan2010}. The regular spacing of the structures could have arisen from a Kelvin-Helmholtz instability driven by shear in the plasma, and thereafter amplified through an interchange instability.

\section{Conclusion and Outlook}
Despite the seeming ubiquity of field-aligned ducts in magnetospheric systems (e.g.~Io-Jupiter \citep{Imai1992} and the Sun \citep{Duncan1979}), self-organization processes in plasmas have been difficult to isolate observationally. Our results demonstrate that widefield radio telescopes such as the MWA are powerful quantitative tools for studying their formation, dynamics and morphology. Radio telescopes differ fundamentally from many approaches for probing the ionosphere and plasmasphere in that they measure density gradients rather than absolute density. Their insensitivity to the constant offset component of the TEC makes them particularly suitable for studying plasma density disturbances/irregularities and permits them to probe regions above the peak electron density with ease, since they are not significantly shielded by the denser underlying plasma. 

The ground-based feature tracking and altitude triangulation capabilities we have demonstrated here offer valuable opportunities for the real-time, regional-scale monitoring of inner magnetospheric structures and dynamics on a near-continuous basis, unconstrained by the limitations of spacecraft orbits or the propagation of whistlers. Unlike ground-based whistler observations, which only allow for radial motions to be measured, the MWA can track both radial and horizontal motions, thereby allowing bulk plasma drifts to be characterized in three dimensions. Its potential to perform 3D reconstruction of density structures can provide empirical constraints on the plasma distribution both along and across magnetic flux tubes. This may be useful, for example, for specifying appropriate boundary conditions in simulation studies of global plasma flows  \citep{Tu2006}.


%
%
%
%
%
%
%

\begin{acknowledgments}
The data supporting this paper are available upon request submitted via email to the corresponding author at sloi5113@uni.sydney.edu.au. This scientific work makes use of the Murchison Radio-astronomy Observatory, operated by CSIRO. We acknowledge the Wajarri Yamatji people as the traditional owners of the Observatory site. Support for the MWA comes from the U.S. National Science Foundation (grants AST-0457585, PHY-0835713, CAREER-0847753, AST-0908884 and AST-1412421), the Australian Research Council (LIEF grants LE0775621 and LE0882938), the U.S. Air Force Office of Scientific Research (grant FA9550-0510247), and the Centre for All-sky Astrophysics (an Australian Research Council Centre of Excellence funded by grant CE110001020). Support is also provided by the Smithsonian Astrophysical Observatory, the MIT School of Science, the Raman Research Institute, the Australian National University, and the Victoria University of Wellington (via grant MED-E1799 from the New Zealand Ministry of Economic Development and an IBM Shared University Research Grant). The Australian Federal government provides additional support via the Commonwealth Scientific and Industrial Research Organisation (CSIRO), National Collaborative Research Infrastructure Strategy, Education Investment Fund, and the Australia India Strategic Research Fund, and Astronomy Australia Limited, under contract to Curtin University. We acknowledge the iVEC Petabyte Data Store, the Initiative in Innovative Computing and the CUDA Center for Excellence sponsored by NVIDIA at Harvard University, and the International Centre for Radio Astronomy Research (ICRAR), a Joint Venture of Curtin University and The University of Western Australia, funded by the Western Australian State government. We thank D.~B.~Melrose for discussions regarding physical interpretations. 
\end{acknowledgments}

\end{article}
%
%
%
%
%
%
%
%


\begin{figure}[H]
  \centering
  \caption{The vector fields of celestial source offsets for the first (a), middle (b) and last (c) snapshots, overplotted with the geomagnetic field lines (black solid lines) from a single $L$-shell. Field line positions were computed based on the Australian Geomagnetic Reference Field model \citep{AGRF2010} at an altitude of 600\,km. Assumed to be locally straight and parallel, their apparent convergence arises from a perspective distortion due to their steep ($-60^\circ$) inclination. The center of each arrow marks the time-averaged position (used as a reference point) of a celestial radio source, and the arrow represents its displacement vector. Arrow lengths have been scaled to 50 times the actual displacement distance. Blue and red denote arrows with positive and negative $x$ components (westward and eastward), respectively. North is up and east is to the left as per the astronomical convention, and $x = y = 0^\circ$ marks the location of the zenith. Units along the $x$ and $y$ axes correspond to angular distance across the MWA FoV ($1^\circ \leftrightarrow$ 10\,km at 600\,km altitude).}
  \label{fig:arrowplots}
\end{figure}

\begin{figure}[H]
  \centering
  \caption{The vector divergence plot for the middle snapshot (i.e.~the divergence of the vector field shown in Fig.~\ref{fig:arrowplots}b), with the duct structure clearly visible. Positive divergence regions (relative overdensities) are represented by red-yellow hues, while negative divergence regions (relative underdensities) are represented by blue-cyan hues. White represents either values close to zero (within the FoV) or NaN (outside the FoV). Geomagnetic field lines have been overplotted as black solid lines, as for Fig.~\ref{fig:arrowplots}.}
  \label{fig:divplot}
\end{figure}

\begin{figure}[H]
  \centering
  \caption{(a) Wavelength ($\lambda$) dependence of the size of the offset for five bright radio sources, whose flux densities are shown in the figure legend (1\,Jy $\equiv 10^{-26}$\,W\,m$^{-2}$\,Hz$^{-1}$). Each set of points represents measurements taken in a single snapshot of time. Straight lines represent best fits to measurements at four different frequencies within the 30-MHz observing bandwidth, constrained to pass through the origin. (b) East-west component of the displacement of one of the sources in (a) as a function of time. Different colors represent measurements at four independent frequencies. Dashed lines indicate the width of the telescope point-spread function. Error bars in (a) and (b) represent position fitting errors. (c) Time dependence of the average magnitude of the TEC gradient over all sources in a snapshot, for images formed using the full 30-MHz band.}
  \label{fig:trends}
\end{figure}

\begin{figure}[H]
  \centering
  \caption{(a) Layout of the 128 MWA tiles, with hollow square and circle markers highlighting the two groups of tiles used for measuring the parallax shift. Central-core tiles (crosses) were excluded to maximize the effective baseline ($\sim$890\,m), leaving 37 tiles for the east half and 35 for the west. (b) Histograms of the parallax shifts measured for the east and west halves, each from 92 independent measurements (46 snapshots $\times$ 2 instrumental polarizations). The relative displacement of the two histograms represents a parallax of $5.3 \pm 0.4$\,arcmin.}
  \label{fig:stereo}
\end{figure}

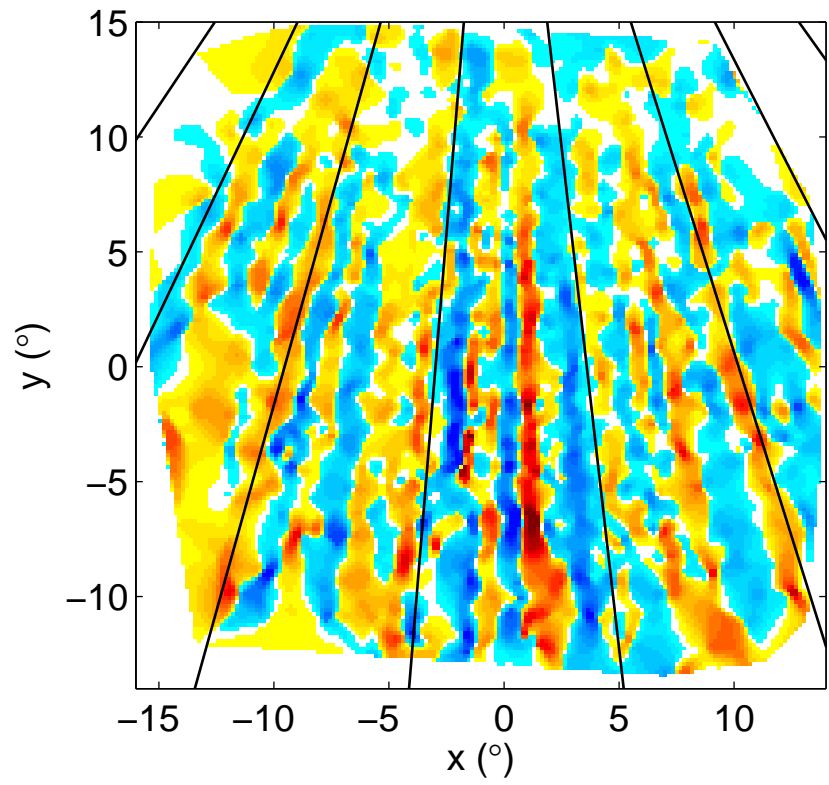
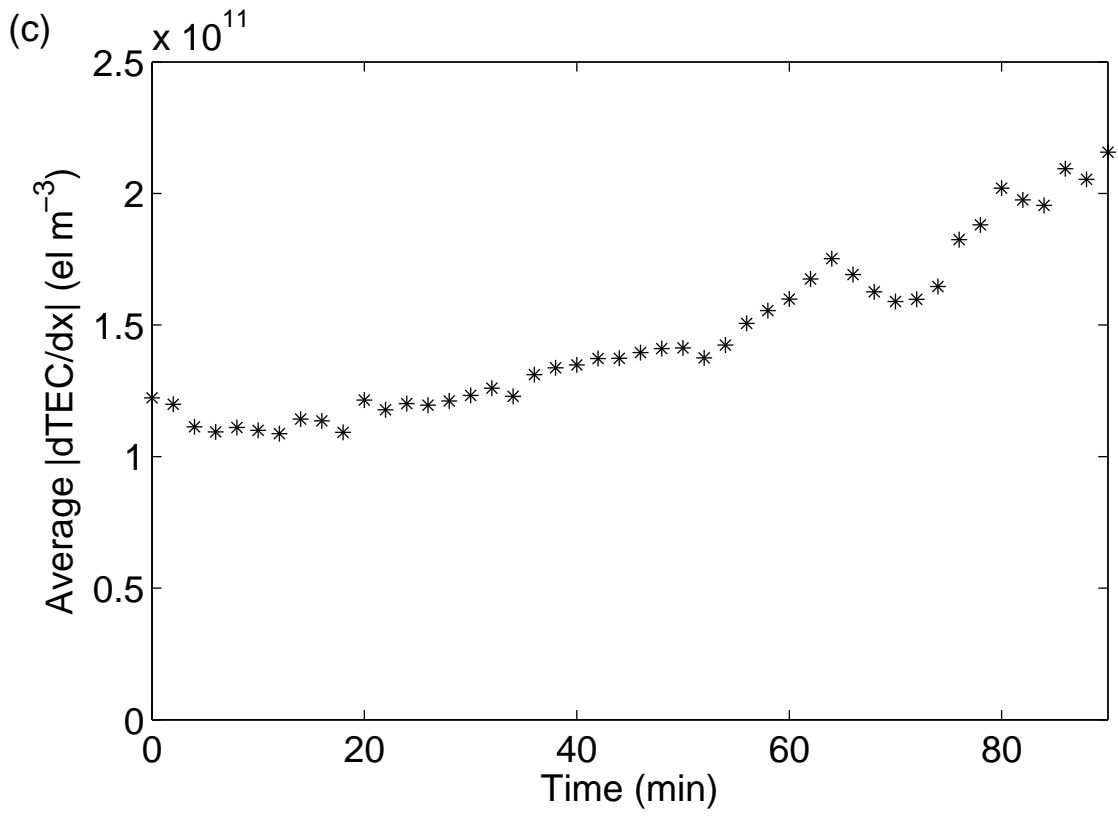
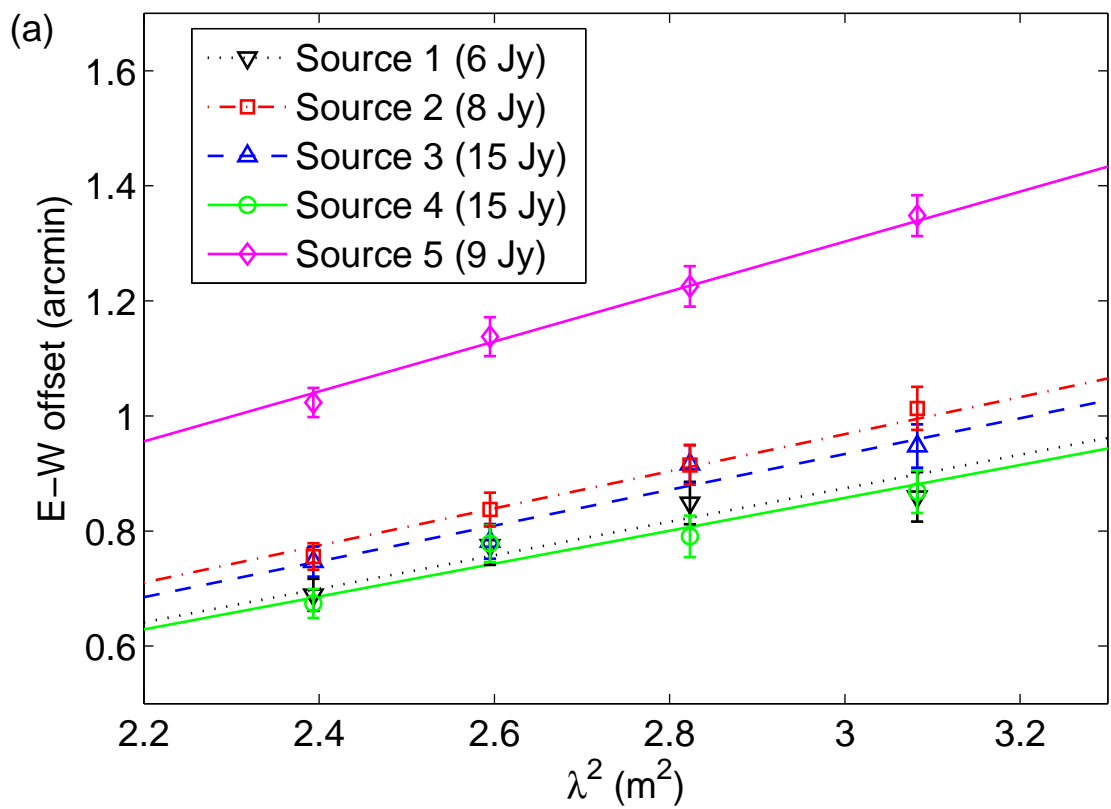
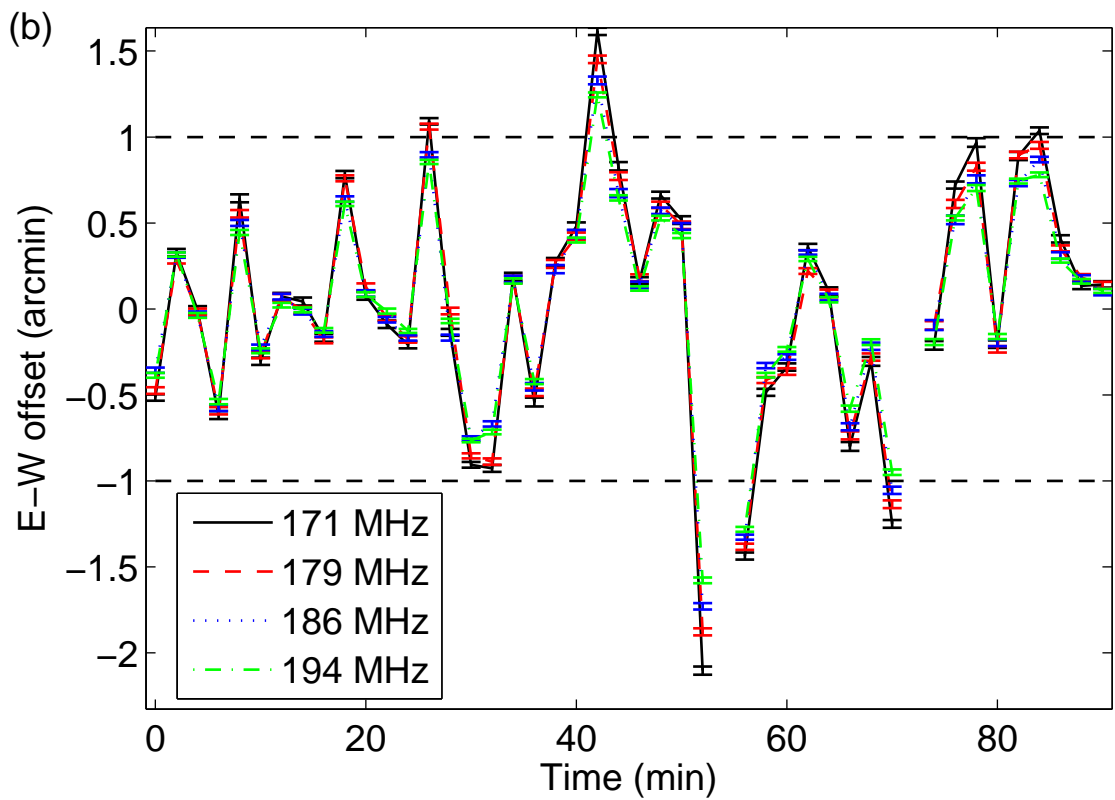
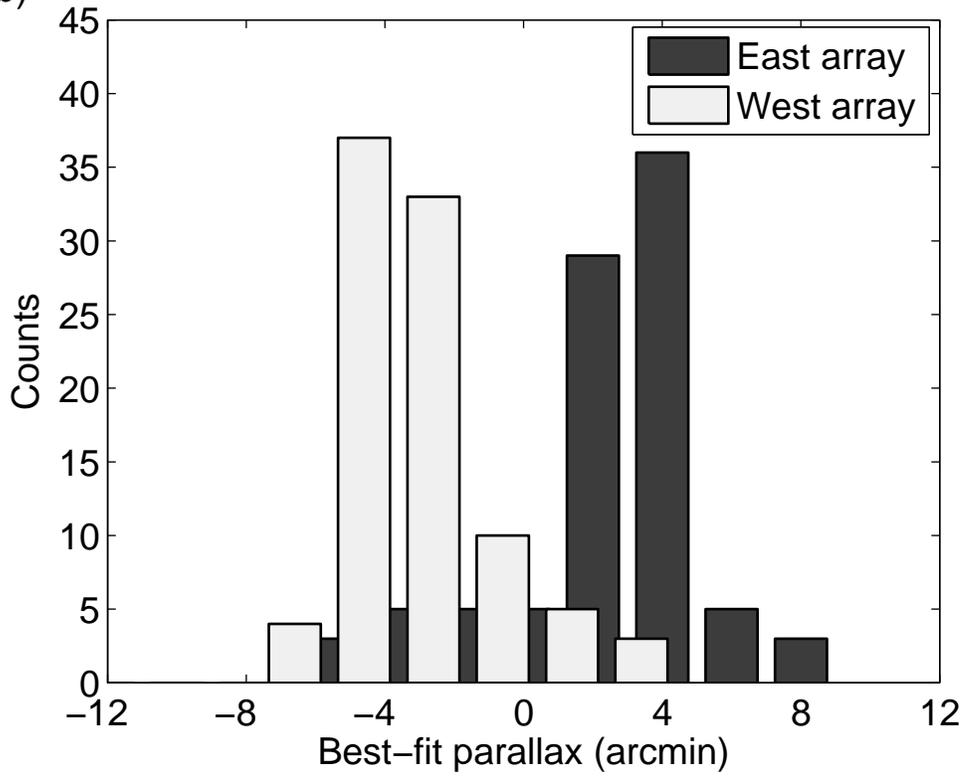
\end{document}